\documentclass[10pt,english,british,tightenlines,eqsecnum,
floats,aps,amsmath,amssymb,nofootinbib,superscriptaddress,prd,showpacs,showkeys]
              {revtex4}

\usepackage[latin9]{inputenc}
\usepackage{color}
\usepackage{babel}
\usepackage{amsmath}
\usepackage{amssymb}
\usepackage{graphicx}
\usepackage{esint}
\usepackage[unicode=true,pdfusetitle,
 bookmarks=true,bookmarksnumbered=false,bookmarksopen=false,
 breaklinks=false,pdfborder={0 0 1},backref=false,colorlinks=false]
 {hyperref}
\makeatletter
\@ifundefined{textcolor}{}
{%
 \definecolor{BLACK}{gray}{0}
 \definecolor{WHITE}{gray}{1}
 \definecolor{RED}{rgb}{1,0,0}
 \definecolor{GREEN}{rgb}{0,1,0}
 \definecolor{BLUE}{rgb}{0,0,1}
 \definecolor{CYAN}{cmyk}{1,0,0,0}
 \definecolor{MAGENTA}{cmyk}{0,1,0,0}
 \definecolor{YELLOW}{cmyk}{0,0,1,0}
}

\@ifundefined{date}{}{\date{25 July 2024}}
\usepackage{babel}

\usepackage{euscript}\usepackage{epsfig}

\usepackage{amsfonts}

\usepackage{fancyhdr}

\makeatother

\usepackage{amsthm}
\theoremstyle{plain}
\newtheorem*{remark*}{Remark}
\usepackage{cleveref}

\begin{document}

\title{Fractional Scalar Field Cosmology}

\author{S. M. M. Rasouli}

\email{mrasouli@ubi.pt}

\affiliation{Departamento de F\'{i}sica,
Centro de Matem\'{a}tica e Aplica\c{c}\~{o}es (CMA-UBI),
Universidade da Beira Interior,
Rua Marqu\^{e}s d'Avila
e Bolama, 6200-001 Covilh\~{a}, Portugal}

\affiliation{Department of Physics, Qazvin Branch, Islamic Azad University,
Qazvin 341851416, Iran}

\author{S. Cheraghchi}

\affiliation{Department of Physics, University of Tehran, P.O. Box 14395-547, Tehran, Iran}

\author{P. V. Moniz}

\email{pmoniz@ubi.pt}

\affiliation{Departamento de F\'{i}sica,
Centro de Matem\'{a}tica e Aplica\c{c}\~{o}es (CMA-UBI),
Universidade da Beira Interior,
Rua Marqu\^{e}s d'Avila
e Bolama, 6200-001 Covilh\~{a}, Portugal}

\begin{abstract}
Considering the Friedmann--Lema\^{i}tre--Robertson--Walker
(FLRW) metric and the Einstein scalar field system as an underlying gravitational model to construct fractional cosmological models has interesting implications in both classical and quantum regimes. %Please check intended meaning has been retained
Regarding the former, we just review the most fundamental approach to establishing an extended cosmological model. We demonstrate that employing new methodologies allows us to obtain exact solutions. Despite the corresponding standard models, we cannot use any arbitrary scalar potentials; instead, it is determined from solving three independent fractional field equations. This article concludes with an overview of a fractional quantum/semi-classical model that provides an inflationary scenario.
 \end{abstract}

\medskip

%\pacs{???????}


\maketitle

\section{Introduction}
\label{SecI}

Many efforts have been undertaken to propose gravitational models so as to overcome the shortcomings of standard cosmology, which was established using general relativity (GR) and the standard model of particle physics. Among these are the monopoles, flatness, horizon, and Big Bang singularity problems.
To address these problems, models including inflationary scenarios and modified gravity frameworks have been proposed
\cite{Linde:1981mu,Cliftona513,brandenberger2000inflationary,faraoni2011early}.

In addition to the previously listed issues, it is worth noting that GR has encountered other shortcomings from both observational and theoretical perspectives that have motivated the application of alternative theories to GR~\cite{Capozziello:2011et,CANTATA:2021ktz,Rasouli:2022tjn,AlvesBatista:2023wqm}.

More precisely, this means that such generalized gravitational frameworks are crucial not only for predicting the observational data and the
achievements of GR  but also for overcoming the shortcomings of existing models~\cite{chokyi2024cosmology,SINGH2024100800}.
Let us be more specific with an example. The cosmological constant has been added to the Einstein--Hilbert (EH) action to address some of the
aforementioned cosmological issues.
Nonetheless, this $\Lambda$CDM model has its own challenges, the most significant of which are the cosmological constant problem~\cite{weinberg1989cosmological,carroll1992cosmological,nobbenhuis2006categorizing,
padilla2015lectures,wang2020reformulation}, Hubble tension problem \cite{Rasouli:2022tjn,dainotti2023hubble,vagnozzi2023seven}, and
fine-tuning and coincident problem~\cite{Sahni:2002kh,Calcagni2017}.

Therefore, to address the problems with the observed discrepancies and limitations of the $\Lambda$CDM model, various generalized models have been proposed, including modified gravity models, such as MOND (Modified Newtonian Dynamics)~\cite{Famaey:2011kh,McGaugh:2014nsa},
$f(R)$ gravity \cite{sharif2014energy,leon2015dynamical,abbas2017higher,sharif2017causes}, entropic gravity~\cite{Padmanabhan:2009vy,verlinde2017emergent}, bimetric gravity~\cite{rosen1973bi,garcia2016cosmology,bassi2023cosmological,maldonado2023axionic,Mousavi:2022puq} minimally coupled Einstein scalar field system~\cite{Rasouli:2019axn,Rasouli:2022tmc,chetia2023particle}; scalar-tensor theories~\cite{Rasouli:2014dba,reyes2018emergence,akarsu2020anisotropic,ildes2023analytic}, specifically, the Brans--Dicke theory~\cite{Rasouli:2016xdo,Rasouli:2016syh,Rasouli:2021xqz}, and their corresponding noncommutative extensions~\cite{Rasouli:2018lny,kan2022classical,Rasouli:2023mae,Rasouli:2022hnp};
generalized Chaplygin gas models~\cite{Bouhmadi-Lopez:2007cag,bouhmadi2015scalar}; and establishing modified fundamental models from their corresponding standard ones in higher dimensions~\cite{Doroud:2009zza,Rasouli:2010zz,Rasouli:2014sda,Rasouli:2017glb,Rasouli:2018owa,Rasouli:2023idg}.

The aforementioned generalized models have various drawbacks despite their many advantages.
In an ongoing effort to address unsolved issues in gravity and cosmology, more potent new models have been proposed in addition to addressing the challenges by modifying the aforementioned models as far as is practical, see, for instance, \cite{RasouliPRD89,Rasouli:2014dba,Rasouli:2016xdo,Rasouli:2016syh,JalalzadehPDU2017}.
 Among the extended cosmological frameworks, \textit{fractional cosmology} is perhaps one of the most promising for revisiting open problems~\cite{roberts2009fractional,Jamil:2011uj,Torres:2020xkw,Gonzalez:2023who}, as demonstrated, for instance, by attempts to address problems like the Hubble tension problem~\cite{Leon:2023}, synchronization problem~\cite{deOliveiraCosta:2023srx}, and cosmological constant problem.
 Still, we think these models need additional presentation, which is why we decided to present this review article.

As we review a fractional cosmological model in both classical and quantum regimes, it is appropriate to provide an overview of how/why we should apply fractional calculus in both regimes.
However, more explanations are presented in the pertinent references.

Fractional calculus is an area of mathematical analysis that extends the differentiation and integration to any real or complex orders.
It can be applied to the study of complex dynamical systems and many other intricate physical phenomena. Regarding its application to cosmology and gravity, it can be considered as extending any gravitational/cosmological framework to non-integer orders in either the classical or quantum regime.
This offers an effective means of investigating gravitational/cosmological phenomena such as the evolution of the universe, the dynamics of black holes, and gravitational waves.

%Most of the study on fractional cosmology in the classical regime
%has focused on important concerns such as A and B.
The majority of classical cosmological models based on fractional derivatives have taken two major approaches~\cite{roberts2009fractional,Shchigolev:2010vh,Shchigolev:2021lbm}:
(i) \textit{Last-step modification:} In this approach, after obtaining the field equations of the model under consideration, ordinary derivatives are replaced by fractional ones.  (ii) \textit{First-step modification:} In this approach, a fractional derivative geometry is established first, followed by the proposal of a fractional-order action~\cite{el2008fractional}.
 It is believed that this method, which can be considered as one of the possible versions of intermediate modification~\cite{roberts2009fractional}, is more fundamental than the previous one.

Regarding fractional quantum physical frameworks, let us provide the following paragraph.
 It is believed that Brownian motion can be considered as the first example of a fractional physical object, where the trajectories (paths) are self-similar, non-differentiable curves whose fractal dimension differs from their topological dimension~\cite{laskin2018fractional,Laskin:2000QM}. Indeed, for the first time, by employing the fractality concept, non-fractional quantum mechanics as a path integral over the Brownian paths was re-established by Feynman and Hibbs~\cite{feynman2010quantum}.
In developing such a formalism, a fractional path integral was then constructed to establish space-fractional quantum mechanics (FQM) as a path integral over the paths of the L\'{e}vy flights, characterized by the L\'{e}vy index $\alpha$, where $0<\alpha\leq 2$, see, for instance,~\cite{Laskin:2000QM}.
In the particular case where $\alpha=2$, the Gaussian process or the process of the Brownian motion is recovered~\cite{Laskin:2000QM}.
An important manifestation of the space FQM is the space-fractional Schr\"{o}dinger equation (SE) in which the second-order spatial derivative of the ordinary SE is replaced properly by a fractional-order derivative (concretely, the quantum Riesz fractional derivative)~\cite{Laskin:2000,Laskin:2000QM}.
Moreover, Naber established the time-fractional SE, in which the first-order time derivative of the ordinary SE was replaced appropriately by a fractional-order derivative (concretely, the Caputo fractional derivative~\cite{almeida2017caputo}),
while the spatial derivative of the ordinary SE remained unchanged~\cite{Naber2004TimeFS}.
Subsequently, the spacetime-fractional SE~\cite{WONG2007,DONG20081005,laskin2017time} was established, in which both the second-order spatial derivative and the first-order time derivative were replaced by the corresponding fractional derivatives as explained above.
In the fractional quantum cosmological framework, we explained how we could retrieve the fractional Wheeler--DeWitt (WDW) equation inspired by the space-fractional SE, using the quantum Riesz fractional derivative for a quantum gravity model~\cite{Rasouli:2021lgy,Rasouli:2022bug}.

The main focus of our work is to review \textit{fractional scalar field cosmology}~\cite{Shchigolev:2010vh,Rasouli:2022bug} briefly and investigate a few cosmological problems. Let us be more specific. We look at the most common cosmological model, in which only a scalar field (minimally coupled to gravity) is added to the EH action, and the universe is described by the FLRW metric.
However, rather than the well-known standard framework, we concentrate on the corresponding lesser-known fractional framework. We discuss such a model in both the classical and quantum regimes. For the former, we first go over a model in which the field equations are obtained from an appropriate fractional action, so that the field equations in a particular case are reduced to standard correspondence.
%This will be followed by brief discussions of the essential concepts from fractional calculus.
Next, we demonstrate how a heuristic approach can be used to analytically obtain the exact solutions to these fractional equations.
We also demonstrate that, in contrast to the standard models, the scalar potential cannot be any ad hoc function of the scalar field.
For the fractional quantum model, we obtain the WDW equation associated with our scalar field cosmology for the non-fractional scenario. Due to the complexities of this equation, we limit our investigation to select circumstances where the exact solution is easily obtained. Then, after a quick discussion of fractional quantum physics, we show how the quantum Riesz fractional derivative can be used to establish our fractional quantum cosmological model. Once again, we look at specific cases as we analyze the fractional WDW equation and show how the results of this fractional model can help us understand the fascinating concepts of non-locality and memory property. In addition to our models herein, the reader may, however, consider other intriguing models like
\cite{calcagni2010quantum,Jamil:2011uj,shchigolev2013fractional,el2017fractional,socorro2023anisotropic} as well as~\cite{el2023paradigm, socorro2023quantum,ccoker2023modified,socorro2023anisotropic,micolta2023revisiting}.

Finally, it is worth noting that the non-locality and memory property of fractional calculus~\cite{du2013measuring,tarasov2018no,tarasov2018generalized,Rasouli:2022bug,tarasov2022nonlocal} have challenged the traditional concepts of locality and Markovian features in classical calculus, motivating scientists to employ fractional calculus broadly in a variety of fields, including physics, and provide a new avenue for a deeper description of various complex systems and phenomena.

Let us conclude our introduction to fractional models here since we go over them in more detail in the following sections at appropriate points.

The structure of the paper is as follows.
A fractional cosmological model corresponding to the Einstein scalar field system in the classical regime is reviewed in the next section.
Then, we present a few novel aspects of the model.
%Moreover, we will briefly go over the mathematical fractional
%foundation before presenting our classical model.
In Section \ref{SecII}, we provide an overview of a fractional quantum model for the same action that is introduced in the next section.
We also explain the fractional calculus required for our quantum model.
For specific cases, the equations for the quantum cosmology model are solved in both non-fractional and fractional cases, revealing intriguing properties and benefits above the corresponding standard models.
Section \ref{Concl} include general discussions about fractional cosmology as well as a brief report on the findings of the presented \mbox{cosmological models.}

%%%%%%%%%%%%%%%%%%%%%%%%%%%%%%%%%%%%%%%%%%%%%%%%%%%%%%%%%%%%%%%%%%
%%%%%%%%%%%%%%%%%%%%%%%%%%%%%%%%%%%%%%%%%%%%%%%%%%%%%%%%%%%%%%%%%%%%%%
%%%%%%%%%%%%%%%%%%%%%%%%%%%%%%%%%%%%%%%%%%%%%%%%%%%%%%%%%%%%%%%%%%%%%%%%
%%%%%%%%%%%%%%%%%%%%%%%%%%%%%%%%%%%%%%%%%%%%%%%%%%%%%%%%%%%%%%%%%%%%%%%%

\section{Classical Perspectives on the Einstein Scalar Field System with the Fractional Action}
\label{Clssical Regime}

As previously noted, the modified gravity models can be established using fractional calculus.
Among the various definitions of fractional integrals and derivatives, the best known are the Riemann--Liouville (RL) and the Caputo~definitions\cite{caputo1967linear,gorenflo1997fractional,mainardi1997fractional}.
To review the fundamentals of fractional calculus, which has been applied to gravity/cosmology, see, for instance~\cite{kimeu2009fractional,herrmann2011fractional,de2014review,baleanu2019fractional,calcagni2021classical,almeida2019variable}.

It has been suggested that the following strategies could be used to modify the conventional cosmological frameworks when using the fractional differential method~\cite{Shchigolev:2010vh}.

 \begin{itemize}

      \item
 \textit{Last-step modification}: this is the simplest approach; the partial derivatives (of the ordinary differential equations associated with the standard cosmological model) are replaced by fractional derivatives.

\item

\textit{Intermediate-step modification}: in this method, instead of employing the standard Lagrangian, the effective field equations are derived from the appropriate fractional Lagrangian, whose terms consist of fractional order derivatives.
 \end{itemize}

 In this section, we only consider the latter approach.
Now, let us re-investigate the fractional scalar field cosmology in the classical regime.
Throughout this work, we consider the FLRW metric
\begin{eqnarray}
\label{FRW-met-1}
{\rm d}s^2 = -N^2(t){\rm d}t^2
+a^2(t)\biggl[\frac{{\rm d}r^2}{1-\mathcal Kr^2}+
r^2({\rm d}\theta ^2+\sin ^2\theta {\rm d}\varphi ^2)\biggr],
\end{eqnarray}
where $N(t)$ is the lapse function, $a(t)$ is the scale factor, and $\mathcal K$ denotes the spatial curvature that takes the values $-1,0,1$ corresponding to the hyperbolic, flat, and spherical three-spaces, respectively.
Moreover, we would like to work with the action
\begin{equation}
\label{act-inf}
S=\int {\rm d}^4x\sqrt{-g}\left[\frac{R}{16\pi G}-\frac{1}{2}g^{\mu\nu}{\nabla}_\mu\phi{\nabla}_\nu\phi
-V(\phi)\right],
\end{equation}
where  we used the units $c=1=\hbar$, and $\phi$ is a homogeneous scalar field minimally coupled to the Ricci scalar $R$. %Please check intended 

In the next section, by considering the action \eqref{act-inf} as an underlying gravitational model, we establish a fractional quantum model.
However, in this section, we focus on the classical model, for which, instead of the standard action \eqref{act-inf}, we apply the fractional integral definition with fractional order $\mu$.
More precisely, the fractional form of \eqref{act-inf} can be considered as~\cite{de2014review}
\begin{eqnarray}\nonumber
 S^{\mu}_{\text{fr}}&=&\int_0^t L^{\mu}_{\text{fr}}d\tau\nonumber\\
&=&\frac{1}{\Gamma(\mu)}\int_0^t Na^3\left\{\frac{3}{8\pi G}\left[\frac{\ddot{a}}{aN^2}+\frac{\mathcal{K}}{a^2}-\frac{1}{N^2}\left(\frac{\dot{a}}{a}\right)^2-\frac{\dot{a}\dot{N}}{a N^3}\right]+\left[\frac{1}{2N^2}{\dot{\phi}}^2-V(\phi)\right]\right\}\\
&\times&(t-\tau)^{\mu-1}d\tau,\nonumber
\label{fr-action}
\end{eqnarray}
where all the functions depend on the intrinsic time $\tau$; an overdot denotes derivative with respect to $\tau$ and we used
\begin{equation}
\label{FRW-Ricci}
R=\frac{6}{N^2}\left[\frac{\ddot{a}}{a}+\left(\frac{\dot{a}}{a}\right)^2-\left(\frac{\dot{N}}{N}\right)\left(\frac{\dot{a}}{a}\right)
+\mathcal K\left(\frac{N}{a}\right)^2\right].
\end{equation}

Employing the Euler--Poisson equation
\begin{equation}
\label{EL}
\frac{\partial L^\mu_{\text{fr}}}{\partial q_i}-\frac{d}{d\tau}\left(\frac{\partial L^\mu_{\text{fr}}}{\partial \dot{q}_i}\right)+\frac{d^2}{d\tau^2}\left(\frac{\partial L^\mu_{\text{fr}}}{\partial \ddot{q}_i}\right)=0,
\end{equation}
and varying the action \eqref{fr-action} with respect to $q_i=\{N,a,\phi\}$, we obtain the equations of motion as
\begin{eqnarray}\label{field eq3}
&H^2+\left(1-\mu\right)\left(\frac{H}{t}\right)=\frac{8\pi G}{3}\rho,\\\nonumber\\
\label{field eq2}
&2\dot{H}+3H^2+2\left(1-\mu\right)\left(\frac{H}{t}\right)+\frac{(1-\mu)(2-\mu)}{t^2}= -8\pi G p,\\\nonumber\\
\label{field eq1}
&\ddot{\phi}+3\left(H+\frac{1-\mu}{3t}\right)\dot{\phi}+\frac{d V(\phi)}{d\phi}=0,
\end{eqnarray}
 where we used the transformation $\tau-t=T\to t$; $H\equiv \dot{a}/{a}$ is the Hubble parameter; and $\rho$ and $p$ stand for the matter associated with the scalar field~\cite{faraoni2004scalar}:
  \begin{eqnarray}\label{rho}
\rho\equiv\frac{\dot{\phi}^2}{2}+V(\phi),\\\nonumber\\
\label{p}
p\equiv\frac{\dot{\phi}^2}{2}-V(\phi).
\end{eqnarray}

   Moreover, we considered only the spatially flat FLRW metric, $\mathcal{K}=0$, and the gauge $N=1$.

 Let us discuss a few aspects of the aforementioned equations.
 \begin{itemize}
     \item
 For $\mu\neq1$, contrary to the standard models,
 %the EMT associated with the scalar field is not conserved, but instead,
 from using the definitions \eqref{rho}, \eqref{p}, and the fractional Klein--Gordon Equation~\eqref{field eq1}, we obtain
 \begin{eqnarray}\label{Con-EMT}
 \dot \rho + 3 H \left(\rho+p\right)=(\mu-1)\left(\frac{\rho+p}{t}\right)\neq 0.
 \end{eqnarray}

 \item
  It is worth noting that in the standard model (where $\mu=1$), there are only two independent equations. However, in this fractional model, we have three independent equations. Namely, we cannot derive the fractional Klein--Gordon (KG) Equation~\eqref{field eq1} from the two others, i.e., Equations \eqref{field eq3} and \eqref{field eq2}.
  \end{itemize}
%%%%%%%%%%%%%%%%%%%%%%%%%%%%%%%%%%%%%%%%%%%%%%%%%%%
%%%%%%%%%%%%%%%%%%%%%%%%%%%%%%%%%%%%%%%%%%%%%%%
%%%%%%%%%%%%%%%%%%%%%%%%%%%%%%%%%%%%%%%%%%%%%%%%%

Assuming $V=0$ and $\mu\neq1$, we can easily show that Equations \eqref{field eq3}--\eqref{field eq1} are inconsistent.
%; such a procedure will be more clear in the next subsection where we will obtain exact solutions for the general case.
As a result, for this particular case, our fractional model differs greatly from the corresponding standard case and the other modified cosmological models that yield interesting results for the vanishing potential, see for instance~\cite{Rasouli:2022hnp}.

Let us use a suitable idea for solving the three independent equations of motion \eqref{field eq3}--\eqref{field eq1}: instead of ${a, \phi, V(\phi)}$, we choose ${H,\rho,p}$ as the unknown variables of the model.
  In this respect, we substitute $\rho$ and  $p$ from \eqref{field eq3} and \eqref{field eq2} into Equation~\eqref{Con-EMT}, which yields a differential equation for $H(t)$ as
\begin{eqnarray}
\label{gen-V-1}
\dot{H}=-3 H^2-\left[\frac{2 (\mu-4 )}{t}\right] H+\frac{(\mu -2) (\mu -1)}{t^2}.
\end{eqnarray}

An exact solution for \eqref{gen-V-1} is
\begin{eqnarray}
\label{gen-V-H}
H(t)=\frac{-2\mu+9-A_{\mu}f(t)}{6 t},
\end{eqnarray}
where
\begin{eqnarray}
\label{gen-V-A-f}
A_{\mu}\equiv\sqrt{16\mu ^2-72 \mu +105},  \hspace{10mm}
f(t)\equiv \frac{1-ct^{A_{\mu}}}{1+c t^{A_{\mu}}}.
\end{eqnarray}
Substituting $H$ from \eqref{gen-V-H} into Equations \eqref{field eq3} and \eqref{field eq2}, we obtain
\begin{eqnarray}
\label{gen-V-rho}
\rho&=&\frac{
\left[A_{\mu} f(t)+8 \mu-15 \right]\left[A_{\mu} f(t)+2\mu -9\right]}{12 t^2},\\\nonumber\\
\label{gen-V-p}
p&=&\frac{\left[A_{\mu} f(t)\right]^2-2(4 \mu-9 ) A_{\mu} f(t)-\left(56 \mu ^2-252 \mu +315\right)}{12 t^2}.
\end{eqnarray}

Moreover, relations \eqref{gen-V-rho} and \eqref{gen-V-p} yield
\begin{eqnarray}
\label{gen-V-phidot}
\dot{\phi}^2&=&\frac{[A_{\mu} f(t)]^2+(\mu -3) A_{\mu} f(t)-5 \mu (4 \mu -15)-90}{6 t^2},\\\nonumber\\
\label{gen-V-V}
V(t)&=&\frac{(3 \mu -7)A_{\mu} f(t)+12 \mu ^2-59 \mu +75}{4 t^2}.
\end{eqnarray}

Now, from using relations \eqref{gen-V-H}, \eqref{gen-V-rho}, and \eqref{gen-V-p}, we
can easily obtain the main unknown variables ${a, \phi, V(\phi)}$.

Let us now persuade ourselves of these exact solutions and postpone a discussion of their relevance to the evolution of the universe.
In future research, the predictions of this model will be compared more closely to those of the relevant standard models or the corresponding more extended cosmological models, as well as to observational data.

%%%%%%%%%%%%%%%%%%%%%%%%%%%%%%%%%%%%%%%%%%%%%%%%%%%%%%%%%%%%%%%%%%%%%%%%%%%%%%
%%%%%%%%%%%%%%%%%%%%%%%%%%%%%%%%%%%%%%%%%%%%%%%%%%%%%%%%%%%%%%%%%%%%%%%%%%%%%%%%%%%
%%%%%%%%%%%%%%%%%%%%%%%%%%%%%%%%%%%%%%%%%%%%%%%%%%%%%%%%%%%%%%%%%%%%%%%%%%%

\section{ADM Formalism and the Fractional Quantum Cosmology}
\label{SecII}

To complete our analysis of the fractional scalar field cosmology, let us study a fractional quantum model in this section. {({It is important to note that the fractional quantum model described here is not the quantized model of the fractional scenario  studied in the preceding section.  Concretely, a different mechanism is utilized to establish it}.)} Due to the complexities of the WDW equation, we confine our investigation to the semi-classical case described in Ref.~\cite{Rasouli:2022bug}.
Therefore, we consider the metric \eqref{FRW-met-1} and the action \eqref{act-inf} as the background geometry and the underlying gravitational model, respectively.

To obtain the Hamiltonian, we begin with the ADM formalism. The corresponding Wheeler--Dewitt (WDW) equation associated with the Friedmann--Lemaitre--Robertson--Walker (FLRW) metric \eqref{FRW-met-1} is then constructed.
 Therefore, we should determine a privileged coordinate to define the velocities. It is almost obvious that slicing the spacetime manifold $\mathcal{M}$ into three-dimensional non-intersecting hypersurfaces with a positive definite metric on it, $(\Sigma,\gamma)$, is a method for separating the coordinates. The metric on $\mathcal{M}$ induces a metric on $\Sigma$ that is denoted by $\gamma_{ij}$. As a result, under this formalism, let us rewrite the FLRW metric as
\begin{eqnarray}
\label{gen-met}
{\rm d}s^2=g_{\mu\nu}{\rm d}x^\mu{\rm d}x^\nu=-N^2(t){\rm d}t^2+
\gamma_{ij}(t){\rm d}x^i{\rm d}x^j.
\end{eqnarray}
%Noting the metric (\ref{FRW-met-1}), it would be convenient to express the time dependency of the induced metric as $\gamma_{ij}=a^2(t)h_{ij}$, where $h_{ij}$ is independent of $t$. Let us assume that the manifold $\mathcal{M}$ is a spatially compact and globally hyperbolic Lorentzian manifold. The volume element in terms of the induced metric on the hypersurface $\Sigma $ is

In Equation (\ref{gen-met}), $\gamma_{ij}=a^2(t)h_{ij}$, where $h_{ij}$ is independent of $t$. Let us assume that the manifold $\mathcal{M}$ is a spatially compact and globally hyperbolic Lorentzian manifold. The volume element in terms of the induced metric on the hypersurface $\Sigma $ is given by
\begin{equation}
\sqrt{-g}~d^4x=N a^3~\sqrt{h}~ d^3x~ dt.
\end{equation}
Moreover, the volume of the spacelike hypersurfaces is
\begin{equation}
\label{defvol}
{\mathcal V}_{{\mathcal K}}\equiv \int {\rm d}^3x\sqrt{h}.
\end{equation}
Furthermore, the scalar curvature and the extrinsic curvature of the hypersurface $(\Sigma, \gamma)$ are
\begin{eqnarray}
    K_{ij}=-\frac{1}{N}\dot\gamma_{ij}=-\frac{a\dot a}{N}h_{ij},\hspace{0.5in}{}^3R=-\frac{6\mathcal{K}}{a^2},
\end{eqnarray}
where a dot denotes the derivative with respect to $t$.
Using the above quantities, the scalar curvature of the manifold $\mathcal{M}$ has the following form
\begin{equation}
    R={}^3R+K^{ij}K_{ij}-K^2+2(\eta^\lambda\eta^\nu_{;\nu}-\eta^n\eta^\lambda_{;\nu})_{;\lambda},
\end{equation}
where ``$;$" denotes the covariant derivative on the manifold $\mathcal{M}$ and
$\eta^\lambda$ is the component of the normal vector of the hypersurface $\Sigma$ satisfying the following relations:
\begin{equation} g_{\mu\nu}X^\mu\eta^\nu=0,\hspace{0.5in}g_{\mu\nu}\eta^\mu\eta^\nu=-1,
\end{equation}
in which  $X$ is an arbitrary vector on the hypersurface $\Sigma$.

Now, considering the aforementioned relations between the geometrical objects on the manifold $\mathcal{M}$ and the hyperspace $\Sigma$, we can express the action of our model in the ADM formalism.

In this section, we reconsider the action \eqref{act-inf}, in which the scalar field is assumed to be a homogeneous field, i.e., it depends solely on $t$.
Using \eqref{act-inf}, \eqref{defvol}, and \eqref{FRW-Ricci}, the ADM action is given by

\begin{equation}
   \label{FRW-action}
S_\text{ADM}=
\int{\rm d}t(L_a+L_\phi)=\int {\rm d}t {\mathcal V}_{{\mathcal K}} Na^3\left[\frac{3}{8\pi G}\left(\frac{\mathcal K}{a^2}-\frac{1}{N^2}\left(\frac{\dot{a}}{a}\right)^2\right)+\left(\frac{1}{2N^2}\dot{\phi }^2
-V(\phi )\right)\right],
\end{equation}
%\end{adjustwidth}
where we eliminated the total derivative term.

The Lagrangian has no $\dot N$ dependence, as observed. Therefore, we consider it a non-dynamical field.

Concretely, we have
\begin{eqnarray}
\label{const}
\frac{\delta S_{ADM}}{\delta \dot{N}}=0.
\end{eqnarray}
The canonical conjugate momenta associated with the dynamical fields $a$ and $\phi$ are
\begin{equation}
\label{momenta}
\Pi _a\equiv\frac{\partial L_a}{\partial \dot a} = -\frac{3{\mathcal V}_{{\mathcal K}}}{4\pi G}\frac{a\dot{a}}{N}, \quad
\Pi _{\phi }\equiv\frac{\partial L_\phi}{\partial \dot\phi} = \frac{a^3{\mathcal V}_{{\mathcal K}}}{N}\dot{\phi }.
\end{equation}
Using the Legendre transform, we can easily obtain the Hamiltonian:
\begin{eqnarray}
H_{ADM}=\Pi_a \dot a+\Pi_{\phi}\dot\phi-L.
\end{eqnarray}
Subsequently, substituting the conjugate momenta into the above Hamiltonian, we finally obtain the canonical ADM Hamiltonian:
\begin{eqnarray}
\label{Hamil1}
H_\text{ADM}= N\biggl[-\frac{2\pi G}{3{\mathcal V}_{{\mathcal K}}a}{\Pi ^2_a}+\frac{1}{{\mathcal V}_{{\mathcal K}}}\frac{\Pi ^2_{\phi }}{2a^3}
-\frac{3{\mathcal K}{\mathcal V}_{{\mathcal K}}}{8\pi G}a +{{\mathcal V}_{{\mathcal K}}}a^3V(\phi )\biggr].
\end{eqnarray}
Using \eqref{const}, we obtain the super-Hamiltonian constrain as
\begin{equation}\label{super}
 \mathcal H=   -\frac{2\pi G}{3{\mathcal V}_{{\mathcal K}}a}{\Pi ^2_a}+\frac{1}{{\mathcal V}_{{\mathcal K}}}\frac{\Pi ^2_{\phi }}{2a^3}
-\frac{3{\mathcal K}{\mathcal V}_{{\mathcal K}}}{8\pi G}a +{{\mathcal V}_{{\mathcal K}}}a^3V(\phi )=0.
\end{equation}
In the following subsections, we analyze the implications of our model for both the standard and the fractional quantum cosmological model.
%%%%%%%%%%%%%%%%%%%%%%%%%%%%%%%%%%%%%%%%%%%%%%%%%
%%%%%%%%%%%%%%%%%%%%%%%%%%%%%%%%%%%%%%%%%%%%%%%%%%%%
%%%%%%%%%%%%%%%%%%%%%%%%%%%%%%%%%%%%%%%%%%%%%%%%%%%%%%%%%%%

\subsection{Wheeler--DeWitt Equation in Slow-Roll Regime}

There are two geometrical structures that can be used to quantize gravity: the diffeomorphism invariant four-geometry of spacetime and the infinite-dimensional symplectic geometry structure (non-degenerate, closed smooth structure) of the space of metrics and momentum on a time-constant slice~\cite{Jalalzadeh:2020bqu}.
One challenge in the quantization of gravity is the method of applying the constraint equations in gravity theory. In general, there are two approaches: the first is to quantize the system and then apply constraints on the wave functions, which is known as the Dirac quantization, and the second is  the reduced phase-space quantization, in which the phase space is obtained by eliminating the constraint equations from the classical equations and then quantizing that space.
Before we quantize our model, let us first describe briefly the concept of the minisuperspace.
The purpose of defining minisuperspace is to use geometrodynamic symmetries to reduce the dynamical system to a finite dimensional system. Given our model's cosmological symmetry, we can introduce the coordinates of the minisuperspace as $q^A=(a(t),\phi(t,x))$, $A=1,2$, such that the metric of this space is defined as
\begin{equation}
f_{AB}=\begin{bmatrix}
-a & 0\\
0 & \frac{4\pi G}{3}a^3\\
\end{bmatrix},
\end{equation}
 where the signature of the metric is $(-,+)$ (for an $N$-dimensional minisuperspace metric, the signature is $(-,+,+,...)$). The action \eqref{FRW-action} in this space is written as
\begin{equation}
S=\int {\rm d}t {\mathcal V}_{{\mathcal K}} N\left[\frac{3}{8\pi G}f_{AB}\dot{q}^A\dot{q}^B-U(q^A)\right],
\end{equation}
where
\begin{equation}
U(q^A)=-\frac{3\mathcal{K}}{8\pi G}q^1-({q^1})^3V(q^2),
\end{equation}
which is employed to indicate the constraint Equation~\eqref{super} in terms of the minisuperspace coordinate:
\begin{equation}\label{minisuper}
\frac{2\pi G}{3\mathcal{V}_\mathcal{K}}f^{AB}\Pi_A\Pi_B+\mathcal{V}_\mathcal{K}U(q^A)=0.
\end{equation}

Now, let us quantize our model by defining the corresponding conjugate momenta as
\begin{equation}
\label{FO}
\Pi _a^2\rightarrow -a^{-p}\frac{\partial }{\partial a}
(a^p\frac{\partial }{\partial a}), \quad
\Pi _{\phi }^2 \rightarrow -\frac{{\partial }^2}{\partial \phi ^2},
\end{equation}
where $p$ is the ordering parameter~\cite{Vilenkin:1987kf} obeying the canonical commutator relations:
\begin{eqnarray}
[a(t), \Pi_a]=i,\hspace{0.5in}[\phi(x,t),\Pi_\phi(y,t)]=i\delta^3(x-y).
\end{eqnarray}
Consequently, equation associated with our cosmological model is given by
\begin{equation}
\label{wdw1}
\frac{{ \partial^2\Psi}}{{\rm \partial}a^2}+
\frac{p}{a}\frac{{ \partial\Psi}}{{ \partial}a}-
\frac{3m_\text{P}^2}{4\pi  a^2}\frac{{\partial^2\Psi}}{{ \partial}\phi ^2}
-\frac{9 {\mathcal V}_{{\mathcal K}}^2 m_\text{P}^4}{16 \pi^2}a^2
\left[{\mathcal K}-\frac{8\pi}{3m_\text{P}^2}{a}^2 V(\phi)\right]\Psi=0,
\end{equation}
where $m_\text{P}=1/\sqrt{G}$ is the Planck mass.
One can easily show that this equation for $p=1$ in terms of the minisuperspace coordinate can be written as
\begin{equation}\label{WDW}
    \left\{\frac{1}{2}\Box+\frac{3\mathcal{V}_\mathcal{K}^2 m_p^2 }{2\pi}U(q^A)\right\}\Psi(q^A)=0.
\end{equation}
In Equation~\eqref{WDW}, $\Box=\frac{1}{\sqrt{-f}}\partial_A(\sqrt{-f}f^{AB}\partial_B)$ is the d'Alembertian operator.
The remainder of this subsection is devoted to solving the WDW Equation~\eqref{WDW} in a specific but crucial case. Our goal in obtaining such a solution is to generate results that may be compared to the impacts of the corresponding fractional case, which are discussed in the following subsection. Let us be more precise.
In what follows, we investigate a slow-roll inflationary scenario associated with the early stages of the universe, which is established based on \eqref{WDW}. Moreover, we restrict ourselves to the following specific conditions.
(i) It should be emphasized that during the slow-roll regime, the inflation potential is a slowly varying function such that we can consider it as a constant. Such a constant potential can be assigned to the  cosmological constant as $ \Lambda\equiv (8\pi V_0)/m_\text{P}^2$~\cite{Linde:1990xn}. (ii) During the slow-roll regime, we can ignore the wave function's $\phi$-dependence. (iii) We  consider only the compact, spatially flat universe, i.e., we set $\mathcal{K}=0$. (iv) We analyze the simplified semi-classical model, i.e., we set $p=0$.
The main reason for the aforementioned simplifications is that, as we see in the following subsection, analyzing the corresponding fractional WDW equation is extremely complicated.

Considering conditions (i)--(iv), the WDW Equation~\eqref{wdw1} reduces to
\begin{eqnarray}
\label{wdw1a}
\frac{{\rm d}^2}{{\rm d}a^2}\Psi (a)+
\frac{9 {\mathcal V}_{{0}}^2 m_\text{P}^4}{16 \pi^2L_0^2}a^4\Psi (a)=0,
\end{eqnarray}
 where $L_0\equiv\sqrt{3/\Lambda}$. {
A solution for Equation~\eqref{wdw1a} is
\begin{eqnarray}
\label{Sol1}
\Psi(a)=\sqrt{a}\left[{\mathcal C}_1 J_\frac{1}{6}\left(\frac{\mathcal V_0 m_P^2}{4\pi L_0}a^3\right)+{\mathcal C}_2 Y_{\frac{1}{6}}\left(\frac{\mathcal V_0m_P^2}{4\pi L_0}a^3\right)\right],
\end{eqnarray}
where $\mathcal{C}_1$ and $\mathcal{C}_2$ are constants. Moreover, $J_\frac{1}{6}$ and $Y_\frac{1}{6}$ denote the Bessel functions (of order one-sixth) of the first and second kind, respectively.}

Considering the sufficiently large value of the scale factor at the present time, the asymptotic behavior of the wave function can help us to retrieve a suitable solution without assigning any specific initial conditions for the universe.
The asymptotic forms of Bessel functions for a large argument can be written as~\cite{malaysia2013derivation}
\begin{eqnarray}
\label{Bes1}
J_\nu(z)\approx\sqrt{\frac{2}{\pi z}}\cos\left(z-\frac{\nu \pi}{2}-\frac{\pi}{4}\right), \hspace{10mm}Y_\nu(z)\approx\sqrt{\frac{2}{\pi z}}\sin\left(z-\frac{\nu \pi}{2}-\frac{\pi}{4}\right).
\end{eqnarray}
In order to determine ${\mathcal C}_1$ and ${\mathcal C}_1$, it is necessary to specify the initial conditions. However, choosing valid initial conditions
 is a complicated procedure, and we do not have enough information about them. Therefore, let us assume ${\mathcal C}_1={\mathcal C}$  and ${\mathcal C}_2=-i{\mathcal C}$ (where ${\mathcal C}$ is a positive real number), which may make a wave function compatible with physical results.
Therefore, assuming  $z\equiv\frac{\mathcal V_0 m_P^2}{4\pi L_0}a^3\gg1 $, $\nu\equiv 1/6$ and employing relations \eqref{Bes1}, the wave function can be written as
\begin{eqnarray}
\label{Bes2}
\Psi(a)={\mathcal C}\sqrt{\frac{8 L_0}{\mathcal V_0 m_P^2}}\exp\left[-i\left(\frac{\mathcal V_0 m_P^2}{4\pi L_0}a^3-\frac{\pi}{3}\right)\right].
\end{eqnarray}

The de Broglie--Bohm interpretation of a quantum mechanical system in quantum cosmology, which is based on the Hamilton--Jacobi formalism, can be considered by assuming the ansatz
\begin{equation}
    \label{SS}
    \Psi(a)=\exp(-iS),
\end{equation}
for the wave function, where $S$ is a function of the scale factor.
 Substituting $\Psi(a)$ from \eqref{SS} into the WDW Equation (\ref{wdw1a}) yields
\begin{equation}
    -\Big|\frac{{\rm d}S}{{\rm d}a}\Big|^2+
\frac{9 {\mathcal V}_{{0}}^2 m_\text{P}^4}{16 \pi^2L_0^2}a^4-i\frac{d^2S}{da^2}=0.
\end{equation}
Assuming $\frac{\mathcal V_0a^3}{4\pi l_P^2 L_0}|\gg1 $, which is equivalent to the WKB condition, $d^2S/da^2\ll (dS/da)^2$, we obtain the Hamilton--Jacobi equation in classical cosmology:
 \begin{equation}
 \label{sh4}
 -\Big|\frac{{\rm d}S}{{\rm d}a}\Big|^2+
\frac{9 {\mathcal V}_{{0}}^2 m_\text{P}^4}{16 \pi^2L_0^2}a^4=0.
\end{equation}
The WKB approximation in quantum cosmology is a semi-classical alternative to considering the wave function. This approach can give a technique to avoid ambiguities caused by operator-ordering concerns in the Wheeler--DeWitt equation and the problems with the path integral formulation of the wave function.
Solving the Equation~\eqref{sh4} gives
\begin{equation}
S=\frac{{\mathcal V}_{{0}} m_\text{P}^2}{4 \pi L_0}a^3+\mathcal{C}_3,
\end{equation}
where $\mathcal{C}_3$ is an integration constant.
Regarding the classical Hamilton--Jacobi theory, the conjugate momentum of the scale factor is given by
\begin{equation}
\label{pa}
 \Pi_a=-dS/da=-\frac{{3\mathcal V}_{{0}} m_\text{P}^2}{4 \pi L_0}a^2.
  \end{equation}
Considering relations (\ref{pa}) and (\ref{momenta}), we can easily obtain $\dot a/a=1/L$ or $a(t)=a(t_0)\exp((t-t_0)/L_0)$. During the inflationary era, where $t_i\leq t\leq t_f$, we conclude
  \begin{equation}\label{sh00}
    \frac{a(t_f)}{a(t_i)}=\exp{\left(\frac{t_f-t_i}{L_0}\right)}.
\end{equation}
This equation shows the expansion rate during the de Sitter expansion.
One can easily see from the above relation that the number of e-folds is $N_e=\ln[a(t_f)/a(t_i)]=\frac{t_f-t_i}{L_0}$. The minimum number of e-folds to solve the Big Bang problems is $N_e\simeq 60$. Moreover, assuming a small time interval for the inflation period such as $10^8t_P$, we obtain $L_0\simeq 10^7 l_p$.

%%%%%%%%%%%%%%%%%%%%%%%%%%%%%%%%%%%%%%%%%%%%%%%%%%%%%%%%%%%%%%%%%%%%%
%%%%%%%%%%%%%%%%%%%%%%%%%%%%%%%%%%%%%%%%%%%%%%%%%%%%%%%%%%%%%%%%%%%%%%%
%%%%%%%%%%%%%%%%%%%%%%%%%%%%%%%%%%%%%%%%%%%%%%%%%%%%%%%%%%%%%%%%%%%%%%

\subsection{Fractional Quantum Cosmology}
\label{FQC}

In this subsection, we use the generalized ADM Hamiltonian and the fractional calculus to establish the corresponding fractional quantum cosmology. Therefore, in what follows, let us present a concise description of the fundamental core of fractional quantum mechanics as well as fractional quantum cosmology.

Laskin was the first to apply fractional derivatives and integrals in quantum mechanics. In the \textit{space-fractional} quantum mechanics he proposed, it was demonstrated that the fractional Hamiltonian was Hermitian and parity-invariant, and he introduced a novel fractional path integral based on the L\'{e}vy flight. Using the space-fractional SE, Laskin calculated the energy levels of the harmonic oscillator and the hydrogen atom~\cite{Laskin:2000,Laskin:2002zz}.
 Furthermore, Naber developed the \textit{time-fractional} Schr\"{o}dinger equation (SE) in another paper~\cite{Naber2004TimeFS}; also see~\cite{Achar2013TimeFS}. This included substituting the Caputo fractional derivative for the first-order time derivative in the standard SE. Dong and Xu~\cite{DONG20081005} and Wang and Xu~\cite{WONG2007} also formulated the \textit{spacetime-fractional} SE in two distinct works.
 In what follows, we limit our discussion to the space-fractional SE.
%The generalized SE is constructed by replacing the time and spatial derivatives with the fractional derivatives.
%\footnote{In SE, all derivative operators will appear as dimensionless objects.}

The ordinary one-dimensional SE associated with the standard non-relativistic quantum physics is given by
\begin{equation}
i\hbar \frac{\partial \Psi(x,t)}{\partial  t}= H\psi(x,t),
\label{ord-SE}
\end{equation}
where (in this section, the hat on the quantum mechanical operators is dropped)
\begin{equation}
H\equiv \frac{p^2}{2m}+V(x),
\label{H-SE}
\end{equation}
is the Hamilton operator in quantum mechanics.
In Equations \eqref{ord-SE} and \eqref{H-SE}, $m$ is the mass of the particle; $x$ and $p=-i\hslash \nabla\equiv-i\hslash \frac{\partial}{\partial x}$ are the quantum mechanical operators; and $V(x)$ and $\Psi(x,t)$ stand for the potential function and the wave function, respectively.

Applying the same method established by Feynman and Hibbs, but from generalizing the Feynman path integral to the L\'{e}vy one (characterized by the L\'{e}vy index $\alpha$, where $0<\alpha\leq2$; letting $\alpha=2$, the Gaussian process is recovered, namely, the L\'{e}vy
motion is replaced by the Brownian motion), the space-fractional SE was obtained~\cite{Laskin:2002zz,Laskin:2000,Laskin:2000QM,Laskin:2000FQM}:
\begin{equation}
i\hbar \frac{\partial \Psi(x,t)}{\partial  t}= D_\alpha
(-\hbar^2 \Delta)^{\alpha/2} \Psi(x,t)
+ V(x) \psi(x,t)\equiv H_\alpha  \Psi(x,t),
\label{SF-SE}
\end{equation}
where $H_\alpha$ is the fractional Hamiltonian operator, $\Delta\equiv \nabla.\nabla$ denotes the Laplacian, $D_\alpha$ is a coefficient with
dimension $[D_\alpha]={\rm erg}^{1-\alpha}{\rm cm}^\alpha{\rm sec}^{-\alpha}$, and $(-\hbar^2 \Delta)^{\alpha/2}$ stands for the
 fractional (quantum) Riesz derivative~\cite{gorenflo1997fractional} in one dimension:
\begin{equation}
(-\hbar^2 \Delta)^{\alpha/2} \psi(x,t)
=
\frac{1}{(2\pi\hbar)}
\int_{-\infty}^{\infty} dp e^{i\frac{p x}{\hbar}}
|p|^\alpha
\int_{-\infty}^{\infty} dx e^{-i\frac{p x}{\hbar}} \Psi(x,t).
\label{RD}
\end{equation}

Another equivalent expression for the fractional Laplacian is:
\begin{equation}
    \label{sh5}
    -\left(-\frac{{\rm d}^2}{{\rm d}a^2}\right)^\frac{\alpha}{2}\Psi(a)=c_{1,\alpha}\int_0^\infty\frac{\Psi(a-v)-2\Psi(a)+\Psi(a+v)}{v^{\alpha+1}}{\rm d}v,
\end{equation}
where $c_{1,\alpha}=\frac{\alpha 2^{\alpha-1}}{\sqrt{\pi}}\frac{\Gamma((1+\alpha)/2)}{\Gamma((2-\alpha)/2)}$~\cite{Pozrikidis_2018}.

It is worth noting that the fractional Laplacian reveals the effects of a non-local process on the conservation law, which is impacted by both local conditions and the general state of the considered field of the model at a certain point in time; for a thorough analysis, see Ref.~\cite{Pozrikidis_2018}.

In the particular case where $\alpha  = 2$ and $D_\alpha = 1/(2m)$, Equation~\eqref{SF-SE} reduces to the standard SE.

From the standpoint of the \textit{phase-space path integral}, it is important to present an outline of fractional quantum mechanics. The fractional SE for the wave function may be obtained from the path integral in the Gaussian case. This formalism, which represents the development of the fractional quantum mechanical system,  modifies the standard SE. The fractional functional measure in the phase-space representation is~\cite{Laskin:2010ry}:
\begin{eqnarray}
&\int_{x_a}^{x_b}\mathcal{D}x(\tau)\int\mathcal{D}p(\tau)=\\\nonumber
&\lim_{N\to\infty}\int_{-\infty}^{+\infty}{dx_1...dx_{N-1}\frac{1}{(2\pi\hbar)^N}}\int_{-\infty}^{+\infty}dp_1...dp_{N}\exp{\left(i\frac{p_1(x_1-x_a)}{\hbar}-i\frac{D_\alpha|p_1|^\alpha\epsilon}{\hbar}\right)}...\times\\\nonumber
&\exp{\left(i\frac{p_1(x_b-x_{N-1})}{\hbar}-i\frac{D_\alpha|p_N|^\alpha\epsilon}{\hbar}\right)}.
\end{eqnarray}
%where $D_\alpha$ is the generalized fractional quantum diffusion coefficien; $1<\alpha\leq2$ is the L\'{e}vi index
The fractional path integral in the continuum limit $N\to\infty$ and $\epsilon\to0$ has the following form
\begin{equation}
K_L(x_b,t_b|x_a,t_a)=\int_{x_a}^{x_b}\mathcal{D}x(\tau)\int\mathcal{D}p(\tau)\exp{\left(\frac{i}{\hbar}\int_{t_a}^{t_b}d\tau[p(\tau)\dot{x}(\tau)-H_\alpha]\right)},
\end{equation}
where $H_\alpha$ is the fractional Hamiltonian given by Equation~\eqref{SF-SE}.
Using the above relations, we can find the evolution of the fractional quantum mechanical system as follows:
\begin{eqnarray}
    \Psi_f(x_b,t_b)=\int_{-\infty}^{+\infty}dx_aK_L(x_b,t_b|x_a,t_a)\Psi_i(x_a,t_a),
\end{eqnarray}
where $\Psi_i$ and $\Psi_f$ are the wave functions at two initial $(t_a)$ and late times $(t_b)$, respectively.
In order to retrieve the fractional SE, the following processes should be conducted:
(i) letting $t_b=t_a+\epsilon$, and expanding the above integral in terms of $\epsilon$, up to the linear order; (ii) using the standard canonical quantization procedure, i.e., $(\mathbf{r},\mathbf{p})\rightarrow (\mathbf{r},\mathbf{-i\hbar\nabla})$, $H\to i\hbar\frac{\partial}{\partial t}$, and then generalizing it to the Riesz fractional derivative.

In what follows, let us present some features of fractional quantum mechanics~\cite{Laskin:2010ry}.
\begin{itemize}
\item {It has been shown that the fractional Hamiltonian $H_\alpha$ is the Hermitian or self-adjoint operator (similar to that of standard quantum mechanics).}

\item For a state of a closed fractional ordinary quantum mechanical system that has a parity, it is straightforward to show that the associated parity is conserved.

\item It is feasible to generalize the fundamental equations of the probability, current density vector, and the velocity vector associated with standard quantum mechanics to retrieve the corresponding ones for fractional ordinary quantum mechanics.
\end{itemize}

To study the applications of the space-fractional SE, see \cite{Padmanabhan:2009vy,verlinde2017emergent}.
It is worth mentioning that both the standard and the space-fractional SE equations obey the Markovian evolution law. To describe the non-Markovian evolution in quantum physics,  the time-fractional SE was formulated \cite{Famaey:2011kh}.   
Using the framework described in this subsection as well as the Hamiltonian formalism derived in the previous subsection, we obtain the fractional WDW equation associated with our cosmological model and attempt to solve it at the semi-classical level.
From the WDW equation, one can find the corresponding wave function and consider the asymptotic behavior of this function.

Let us now present an overview of our fractional quantum cosmological model. The fractional ADM Hamiltonian is obtained by employing the transformation
\begin{equation}
\frac{2\pi G}{3\mathcal{V}_\mathcal{K}}f^{AB}\Pi_A\Pi_B\to \frac{2\pi}{3\mathcal{V}_\mathcal{K}m_p^\alpha}\left(f^{AB}\Pi_A\Pi_B\right)^{\frac{\alpha}{2}}
\end{equation}
into Equation~\eqref{minisuper}, which provides
\begin{equation}\label{FH constraint}
H^\alpha_{ADM}=\frac{2\pi}{3\mathcal{V}_\mathcal{K}m_p^\alpha}\left(f^{AB}\Pi_A\Pi_B\right)^{\frac{\alpha}{2}}+\mathcal{V}_\mathcal{K}U(q^A)=0.
\end{equation}
Quantizing the fields and using the constraint equation lead to the fractional WDW equation
\cite{Jalalzadeh:2020bqu}:
\begin{equation}\label{WDWsh}
    \left\{\frac{m_P^{2-\alpha}}{2}(-\Box)^\frac{\alpha}{2}-\frac{3\mathcal{V}_\mathcal{K}^2 m_p^2 }{2\pi}U(q^A)\right\}\Psi(q^A)=0,
\end{equation}
where we have used
\begin{equation}\label{x}
%    \begin{array}{cc}
(-\Box)^{\alpha/2} \Psi(q^A)
=\mathcal F^{-1}(|\mathbf{p}|^\alpha(\mathcal F\Psi(\mathbf p)).
%\end{array}
\end{equation}
In Equation~\eqref{x}, $|\mathbf p|=\sqrt{\Pi^A\Pi_A}$ and $\mathcal F$ denotes a Fourier transformation.
We discussed a slow-roll inflationary scenario in the preceding subsection. Applying the same assumption to the fractional model in \eqref{WDWsh}, we obtain the simplified WDW equation as
\begin{eqnarray}
\label{FWDW}
-\left(-\frac{{\rm d}^2}{{\rm d}a^2}\right)^\frac{\alpha}{2}\Psi (a)+
\frac{9 {\mathcal V}_{0}^2 m_\text{P}^{\alpha+2}}{16 \pi^2L^2}a^4\Psi (a)=0.
\end{eqnarray}
It should be noted that in the case where $\alpha=2$, Equation~\eqref{FWDW} reduces to its standard counterpart \eqref{wdw1a}.
To solve the fractional WDW equation in the semi-classical limit, we employ the exponential form of the wave function, i.e., \eqref{SS}, in Equation~\eqref{FWDW}, then compare the equation with \eqref{sh5}, expand $\Psi(a\pm v)$ in a Taylor series about $a$, and finally derive the fractional WDW equation using the WKB approximation:
\begin{equation}
    \label{sh6}
    -\left(-\frac{{\rm d}^2}{{\rm d}a^2}\right)^\frac{\alpha}{2}\Psi(a)=c_{1,\alpha}e^{-iS(a)}\int_0^\infty\frac{\sin^2(\frac{vS'}{2})}{v^{\alpha+1}}dv=|S'|^\alpha \Psi(a),
\end{equation}
where $S'\equiv dS/da$. The semi-classical solution of this equation up to a pre-exponential factor is:
\begin{equation}
    \label{sh9}
    \Psi(a)\propto\exp{\left\{i\left( \frac{3\mathcal V_0m_P^{\frac{\alpha}{2}+1}}{4\pi L}\right)^\frac{2}{\alpha}a^\frac{4+\alpha}{\alpha}\right \}}.
\end{equation}
Employing the fractional Hamiltonian constraint \eqref{FH constraint} and the assumptions of our model gives
\begin{equation}
    \label{sh7}
    -\frac{2\pi}{3{\mathcal V_0}m_P^\alpha a}|\Pi_a|^\alpha+
    \frac{3 {\mathcal V}_{0} m_\text{P}^{2}}{8 \pi L^2}a^3=0,
\end{equation}
\begin{equation}
    \label{sh8}
    H_\text{ADM}^{(\alpha)}=N\left[ -\frac{2\pi}{3{\mathcal V_0}m_P^\alpha a}|\Pi_a|^\alpha
    + \frac{3 {\mathcal V}_{0} m_\text{P}^{2}}{8 \pi L^2}a^3\right].
\end{equation}
From the above relations, one can easily retrieve Hamilton's equations:
\begin{equation}
    \label{sh10}
    \dot a=\frac{\partial H^{(\alpha)}_\text{ADM}}{\partial \Pi_a},~~~ \dot \Pi_a=-\frac{\partial H^{(\alpha)}_\text{ADM}}{\partial a}.
\end{equation}
Finally, using the Hamiltonian constraint (\ref{sh7}) and letting $N=1$, we find the scale factor:
\begin{equation}
    \label{sh11}
    a(t)=\left[\frac{2(D-2)}{(D-1)} \right]^\frac{1}{2(D-2)}\left(\frac{4\pi L}{3\mathcal V_0m_P}\right)^\frac{1}{2}\left(\frac{t}{L}\right)^\frac{1}{2(D-2)},~~~~~2<D<3,
\end{equation}
where $D\equiv2/\alpha+1$~\cite{Rasouli:2022bug}.

Let us now describe the evolution of the universe according to relation \eqref{sh11}, which is obtained by employing a simple fractional quantum cosmological scenario.\\

    A scalar field with an appropriate potential and dominating the early energy density of the universe is typically employed to achieve inflation, causing the scale factor of the universe to expand rapidly. In the original inflation theories, this expansion is exponential and resembles the de Sitter space as the scalar field progressively rolls down to its global minimum. However, in our fractional modification of the same model, we obtain the deceleration parameter as
    %Therefore, during inflation, the deceleration parameter is $q=-1$. However, in fractional modification of the same model, we obtain
    \begin{equation}
        \label{sh12}
        q=-\frac{\ddot a/a}{H^2}=2D-5,
    \end{equation}
    where we used (\ref{sh11}).
   
The early universe accelerates if $2<D<2.5$, as predicted by Equation~\eqref{sh12}, and for these values of $D$, we have a power-law inflationary model. As seen, the L\'evy's fractional parameter $\alpha$ plays a crucial role in this model.

    %\item
    Assuming that the inflationary epoch occurred during a time interval $t_f-t_i=10^{-37}~s=10^8t_P$ after Planck's time, we obtain the number of e-folds as
    \begin{equation}
        \label{sh13}
        N_e=\ln\left(\frac{a(t_f)}{a(t_i)}\right)=\frac{1}{2(D-2)}\ln\left(\frac{t_f}{t_i}\right)=\frac{\ln(10)}{2(D-2)}\simeq\frac{1}{D-2},
    \end{equation}
where we used the fractional form of the scale factor, i.e., Equation (\ref{sh11}).
As an example, if $D=2.019$, we obtain $N_e\simeq60$, which is the smallest number of e-folds required to solve the standard problems of the Big Bang cosmology.

\section{Conclusions and Discussions}
\label{Concl}
In this short review article, we studied the application of fractional calculus in classical and quantum cosmology.
More concretely, considering the Einstein scalar field system and the FLRW metric, we established fractional frameworks to study the corresponding cosmological problems in classical and quantum regimes.

  We presented the necessary foundations of fractional calculus for both regimes and discussed potential fractional quantum physics and fractional cosmological frameworks. As was noted in the preceding sections, the classical cosmological models can be generalized to two different methods when employing fractional calculus. In addition, three approaches to extend standard quantum physics were presented. We additionally provided a brief explanation of the derivation of the fractional SE, and  how an intriguing fractional quantum cosmology could be established inspired by fractional quantum mechanics. In summary, this review article aimed to draw attention to the fascinating fractional approaches found in quantum physics, gravity, and cosmology. To achieve this, it basically outlined relevant interesting models and introduced books, articles, and their authors in order to maximize attention.

Let us provide a bit more details now.
In the classical regime, we reviewed a cosmological model that considered the spatially flat FLRW metric and merely a scalar field minimally coupled to the Ricci scalar, with no ordinary matter included. We then concentrated on a fractional  modification approach to obtain the fractional field equations. It is critical to note that, unlike the corresponding conventional model, this model had three independent equations for three unknowns $a$, $\phi$, and $V(\phi)$. It should be noted that in the standard model, there are only two independent equations to which the scalar potential is considered through ad hoc assumptions or the consideration of specific conditions. However, in our fractional model, the potential was also an independent variable that could not be assumed to be any arbitrary function but was obtained exclusively by solving the field equations, requiring that all three equations be consistent. Assuming, for instance, that the potential was zero, we demonstrated that the fractional equations became inconsistent, whilst intriguing solutions were obtained in the corresponding standard model or its generalized versions (see, for instance,~\cite{Rasouli:2022hnp}).
Subsequently, we solved the field equations in the general case without taking into account any limiting conditions, which was made feasible by a novel approach: we assumed that $\{H,\rho,p\}$ were new unknowns in place of the previously specified unknowns, which allowed us to find exact analytical solutions.

To wrap up our study of the fractional scalar field cosmology, we attempted to present a coherent description of it in the quantum regime. Let us be more precise. Considering the FLRW metric together with the Einstein scalar field system and using the Hamiltonian approach, we obtained the WDW equation in a completely general form. However, because obtaining exact solutions analytically was unfeasible, we limited our analysis to a specific but most significant case. First, we obtained the solutions for the standard WDW equation by considering the WKB condition, slow-roll regime, and $p=0$, where $p$ was the ordering parameter. Our findings showed that the scale factor evolved exponentially in early periods, coinciding with the de Sitter expansion. Then, we showed that under certain conditions, the problems with the standard cosmology could be solved in that model.

 After that, we gave a summary of a fractional quantum cosmological model. More precisely, we demonstrated how to apply the same procedure that yielded the space-fractional SE to construct the fractional WDW equation. To solve this fractional WDW equation, we had to apply the aforementioned assumptions to the corresponding differential equation once more. We found that the scale factor behaved significantly differently than in the standard case. Concretely, we showed that the scale factor was a power-law function of time that depended on the volume of the compact three-space and the L\'{e}vy fractional parameter.
In~\cite{Rasouli:2022bug}, it was stressed that such interesting results,
%on new parameters,
along with the distinctive features of fractional calculus in cosmology, their interpretation, and the reasons behind their basic distinctions from classical models, were all highly debatable.  The models reviewed in this article can considerably increase our motivation to propose new fractional models, which may hold the key to re-examining outstanding problems in cosmology.\\
\\

Finally, it is worth noting a few important points:
\begin{itemize}
\item {As has been highlighted in the preceding sections, this study is merely a brief \textit{review article} on the topic, and our goal in writing this article was to provide a review of the few prior studies in a coherent and abstract manner as well as to add a few comments to emphasize the importance of fractional cosmology established by considering the most commonly used Einstein scalar field system.}

 \item {
 It should be noted that the scenario reviewed in Section \ref{FQC} is not the corresponding quantized scenario of the classical one reviewed in Section \ref{Clssical Regime} (that is why two distinct fractional parameters were utilized).}

 \item {
 We should point out that, in this article, we only looked at the specific fractional cosmological models in the particular cases especially for the quantum regime.
More precisely, the latter can be investigated for more extended cosmological models such as those with either general scalar potentials, non-vanishing spatial curvature, or non-vanishing factor-ordering parameter, see, for instance~\cite{Coule:1999wg,Socorro:2009rd} and references therein.}

\item {Our models can be extended to more generalized models by replacing either the FLRW metric or the Einstein scalar field setting by other metrics or more extended underlying gravitational models.
Concretely, by considering each of the standard cosmological models, it is possible to establish the corresponding fractional model and compare the obtained results not only to those of the standard models but also to determine whether such results can accurately predict the reported recent observational data.
In any case, fractional cosmology is a relatively new and powerful paradigm that has yet to be applied to outstanding problems in cosmology.
Our future efforts will include producing other interesting models within this scope.}

 \end{itemize}

\section*{Acknowledgments}
We express our gratitude to the reviewers for their insightful feedback.
SMMR an PM acknowledge the FCT grants
 UID-B-MAT/00212/2020 and UID-P-MAT/00212/2020
at CMA-UBI plus the COST Action CA18108 (Quantum gravity phenomenology in the multi-messenger approach).

\bibliographystyle{utphys}

\end{document}